\documentclass[graybox,natbib,nosecnum]{svmult}
\bibpunct{(}{)}{;}{a}{}{,} %

\pdfoutput=1   %

\usepackage{amsmath,amssymb}
\usepackage{mathptmx}       %
\usepackage{helvet}         %
\usepackage{courier}        %
\usepackage{type1cm}        %

\usepackage{makeidx}         %
\usepackage{graphicx}        %
\usepackage{multicol}        %
\usepackage[bottom]{footmisc}%
\usepackage[normalem]{ulem}	%
\usepackage[dvipsnames]{xcolor}
\usepackage[colorlinks=true,linkcolor=MidnightBlue,citecolor=MidnightBlue,urlcolor=MidnightBlue]{hyperref}
\usepackage{soul}   %

\newcommand{\hbindex}[1]{{#1}\index{#1}}  %

\makeindex             %

\begin{document}

\title*{Dynamical Evolution of Planetary Systems}
\author{Antoine C. Petit, Gabriele Pichierri, Max Goldberg and Alessandro Morbidelli}
\authorrunning{A.
Petit, G.
Pichierri, M.
Goldberg, A.
Morbidelli} %
\institute{Antoine  C. Petit \at Universit\'e C\^ote d'Azur, CNRS, Observatoire de la C\^ote d'Azur, Laboratoire J.-L.Lagrange, Nice, France \\\email{antoine.petit@oca.eu}
\and
Gabriele Pichierri \at Dipartimento di Fisica, Università degli Studi di Milano, Milano, Italy
\at California Institute of Technology, Pasadena, USA
\and
Max Goldberg \at Universit\'e C\^ote d'Azur, CNRS, Observatoire de la C\^ote d'Azur, Nice, France
\and
Alessandro Morbidelli
\at Collège de France, CNRS, PSL Univ., Sorbonne Univ., Paris, France
\at Universit\'e C\^ote d'Azur, CNRS, Observatoire de la C\^ote d'Azur, Nice, France
}
\maketitle

\abstract{Planetary systems can evolve dynamically even after the planets themselves have fully formed, and there is circumstantial evidence that most planetary systems become unstable after the disappearance of the gaseous protoplanetary disk. 
Theories of planet formation predict that chains of mean motion resonances are the natural outcome of disk-driven planet migration, leading to the pile up of super-Earths resonant chains close to the inner edge of the disk and the formation of fragile chains for distant giant planets. 
Observations of young systems suggest that they are more often locked in these chains than older ones, which are instead mostly non-resonant. The instabilities thought responsible for this trend can arise intrinsically if the original systems are too closely packed, or be due to external perturbations such as tides, planetesimal scattering, or torques from distant stellar companions. The Solar System was not exceptional in this sense, as the outer giants saw the disruption of a resonant chain; meanwhile, the inner system was likely built through a series of giant impacts between closely packed planetary embryos. Thus, the orbital distributions of planetary systems that is observed today, both solar and extrasolar, can be different from those emerging from formation and assembly processes within the disk, and it is important to consider possible long-term dynamics to connect the two.
}

\section{Introduction }
\label{sec:intro}

This chapter concerns the \hbindex{dynamical evolution} of planetary systems after the removal of gas from the protoplanetary disk.
Most massive planets are expected to form within the lifetime of the gas component of protoplanetary disks 
and, while they form, they are expected to evolve dynamically due to gravitational interactions with the gas 
However, the dynamical evolution of planetary systems is not over once the gas disappears.
The next section will review the range of observational evidence for post-gas evolution.
Most planetary systems that are observed are around main-sequence stars and therefore this post-gas evolution needs to be taken into account if one wants to explain their current orbital characteristics.
This chapter continues by discussing planetary systems becoming unstable by themselves, i.e. without interactions with planetesimals or stellar companions, and the outcome of these instabilities.
Finally, the external processes that may destabilize or otherwise modify systems are covered.
These include interactions with the disappearing gas-disk, with a remnant planetesimal disk, with a stellar companion, or tidal interactions with the host star.

\section{Observational evidence for post-gas evolution of planetary systems}
\label{sec:observations}

The Solar System holds the most direct evidence of post-gas dynamical evolution. The asteroid belt, the Kuiper belt and the Trojans of Jupiter and Neptune all have excited eccentricities and inclinations, which would have been damped by the gas \citep{Adachi1976, Brasser2007} and thus were attained after the disk dissipated. Moreover, their orbits are stable with today's orbital configuration of the giant planets, indicating that the planets' orbits were originally different and evolved to the current configuration sometime in the post-gas era of the Solar System. Indeed, planet migration theory predicts that the giant planets should have migrated convergently and locked in mutual mean motion resonances, wherein the ratios of orbital periods of the planets are close to ratios of small integers \citep{Morbidelli2007}. The giant planets thus must have evolved from this primordial multi-resonant configuration to the current non-resonant one.
They could have done so only after gas removal; otherwise gas-driven migration would have brought them back into resonance.
The so-called Nice model \citep{Morbidelli2007,Levison2007,Levison2011,Batygin2012a,Nesvorny2012,Griveaud2024} proposes that this happened during a phase of dynamical instability and shows how this instability sculpted the small body populations.

\begin{figure}[t]
    \centering
    \includegraphics[width=0.49\linewidth]{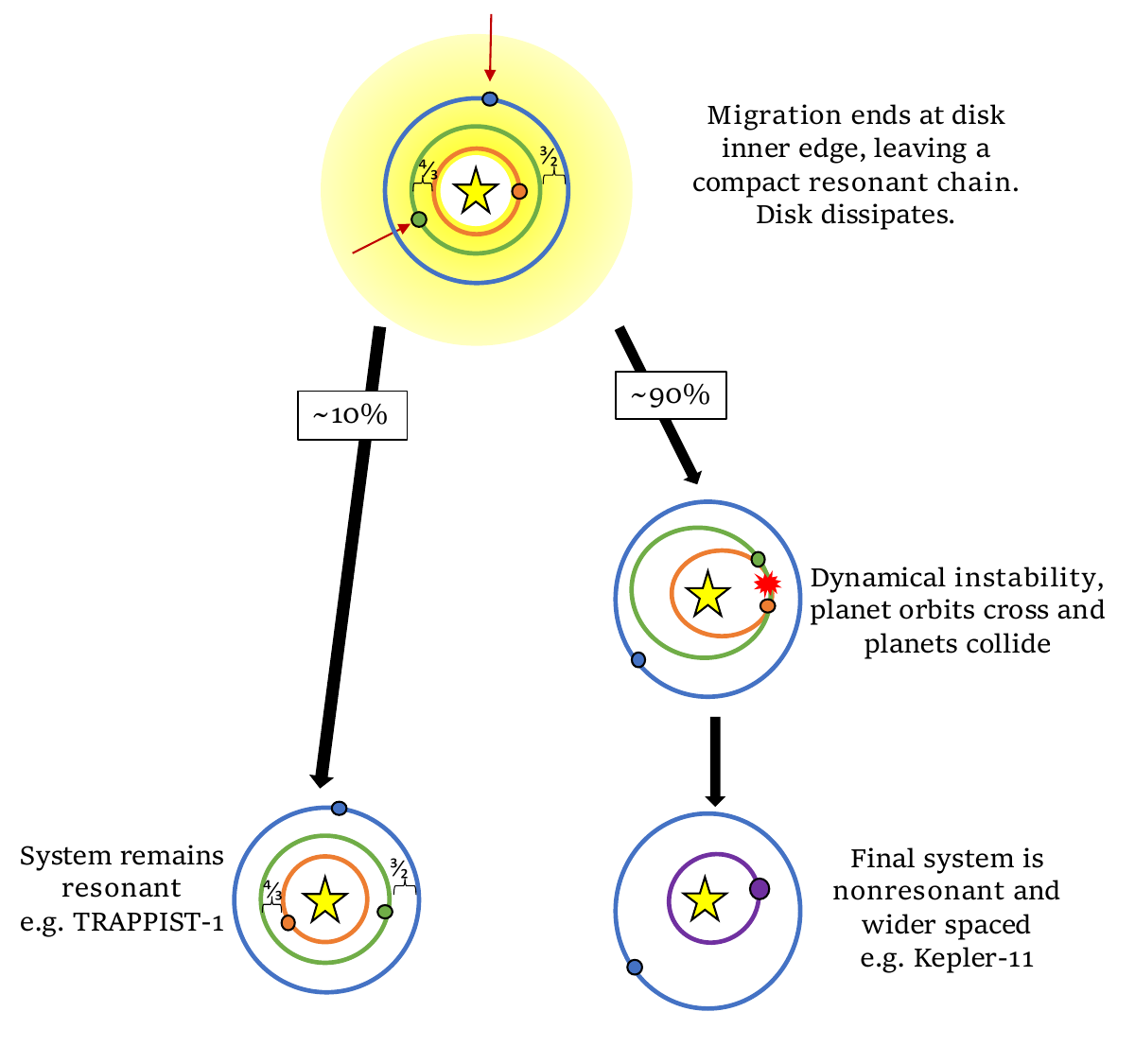}
    \includegraphics[width=0.49\linewidth]{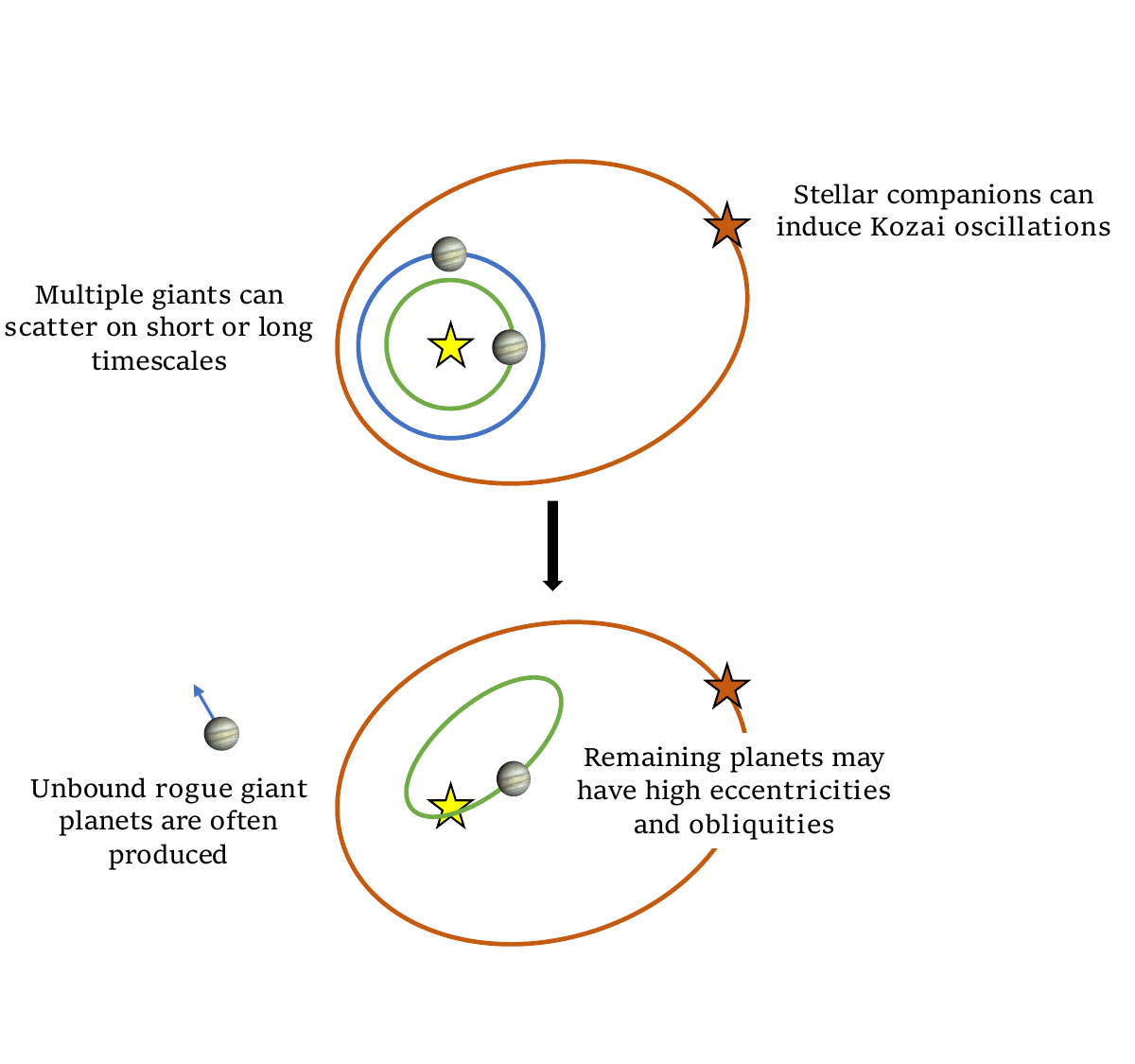}
    \caption{\emph{Left panel:} Skecth decribing the fate of compact close-in exoplanet systems. Planets pile up near the inner edge of the disk in resonant chains that are in their majority broken once the disk dissipates.
    \emph{Right panel:} Typical outcome of the scattering phase for distant giant planets. Under self or induced destabilization, one or several planet can be ejected, leaving the remaining planets on very eccentric and inclined orbits.}
    \label{fig:cartoon}
\end{figure}

Extrasolar planets provide evidence that orbital reconfiguration after gas removal is the norm rather than an exception (Fig.~\ref{fig:cartoon}). 
Extrasolar giant planets are frequently found on eccentric and inclined orbits.
The best explanation for the statistical distribution of their orbital parameters is that the observed planets were initially part of a multi giant-planet system which became violently unstable, some of the original planets being ejected or tossed onto undetectable long-period orbits \citep{Ford2008,Juric2008,Chatterjee2008,Beauge2012,Pu2021,Li2021}.
As in the Solar System, these instabilities should have occurred after gas dissipation because otherwise the \hbindex{planet-disk interactions} would have re-stabilized the planetary orbits in a new compact configuration with low-eccentricities \citep{Lega2013}.

In contrast with extrasolar giant planets, super-Earths typically have less dynamically excited orbits \citep{VanEylen2019}.
Nevertheless, substantial evidence suggests that the vast majority of super-Earth systems (probably more than 90\%) experienced post-gas instabilities as well \citep{Izidoro2017,Izidoro2021}.
Firstly, super-Earths should have been in multi-resonant configurations at the end of the gas-disk lifetime as a consequence of migration, while the observed distribution of orbital periods of adjacent super-Earths shows a broad distribution with no preference for ratios of small integers (i.e. resonances).
In fact, the typical spacing between super-Earths tends to be around 20 mutual Hill-radii (see the next section for a more in-depth discussion), whatever period ratio that implies \citep{Pu2015}. 
Only a few systems are confirmed to be in resonant configurations \citep[e.g.][]{Mills2016,Agol2021,Leleu2021}.
Indeed, \hbindex{resonances} appear to be prevalent among the youngest compact systems, suggesting that the dynamical reorganization takes place over several 100 Myrs \citep{Dai2024}.

\begin{figure}

\includegraphics[width=0.31\linewidth]{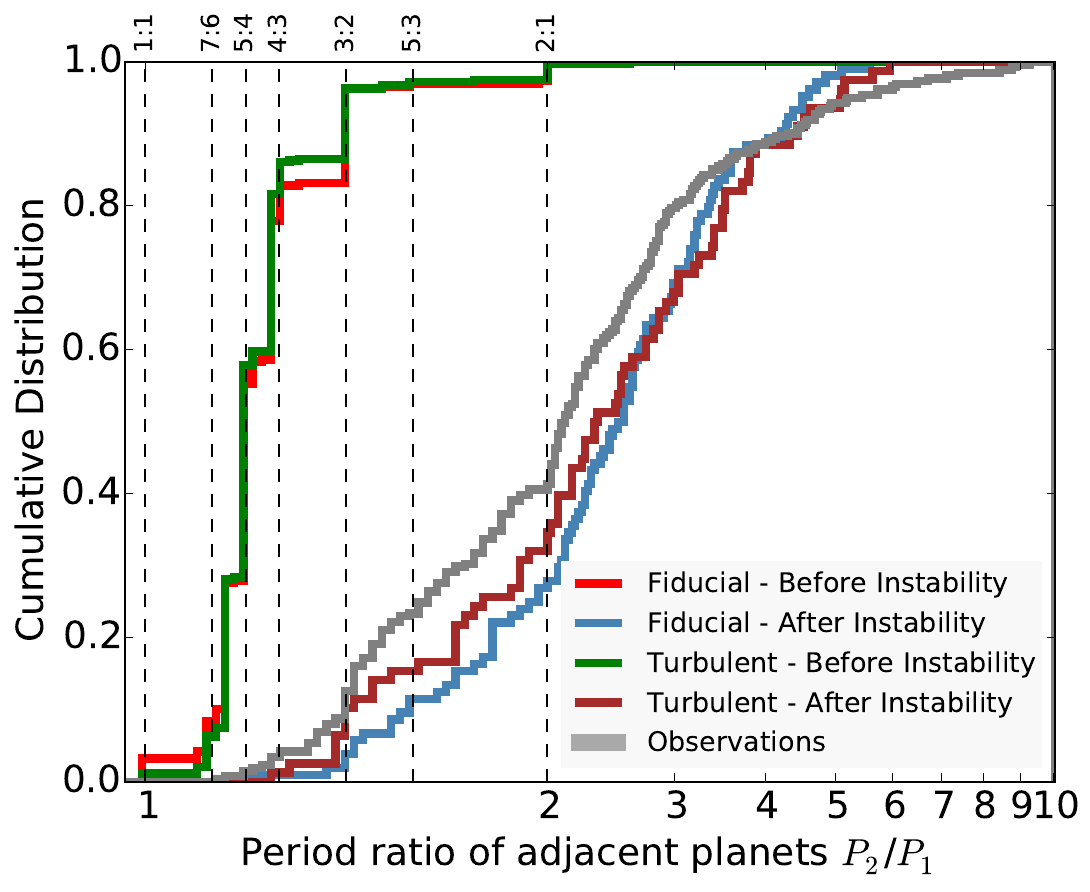}\,\includegraphics[width=0.31\linewidth]{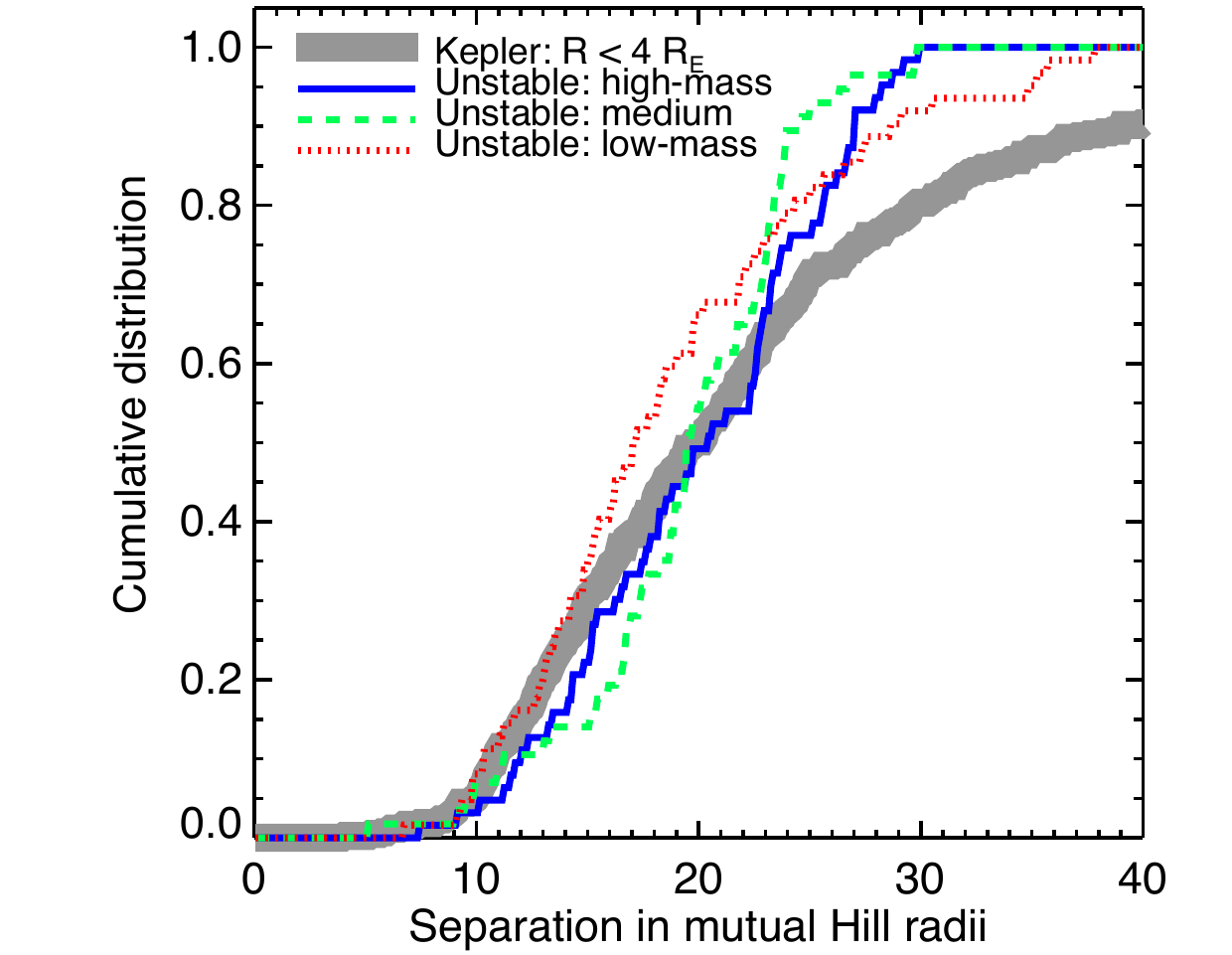}\,\includegraphics[width=0.37\linewidth]{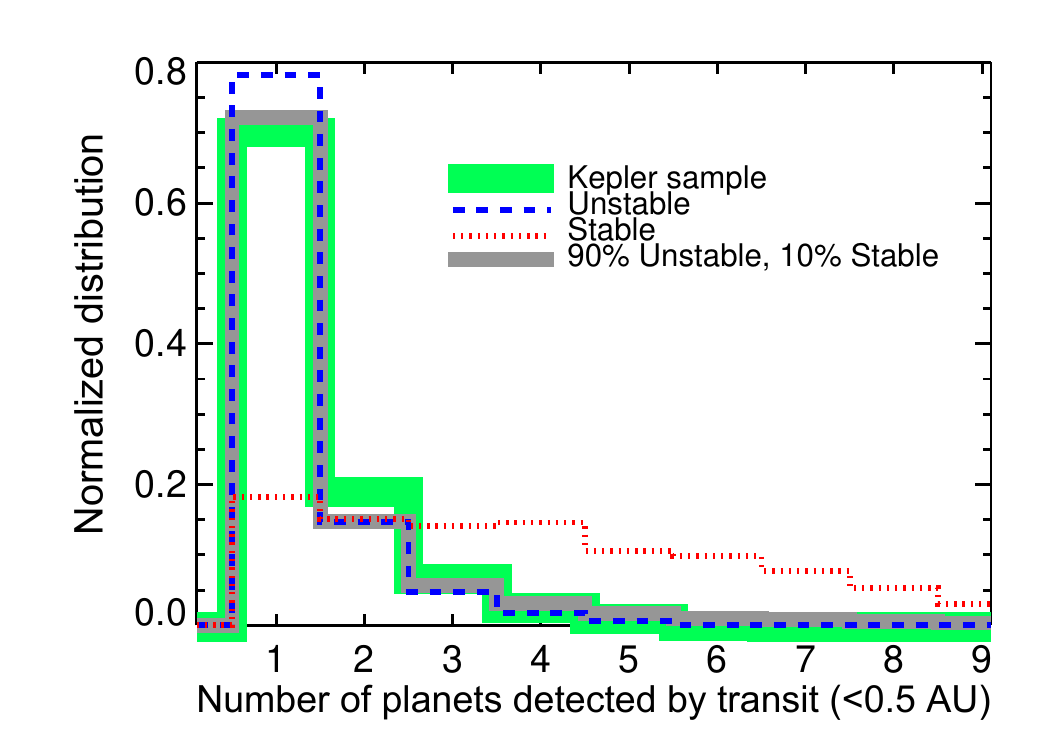} 
\caption{\emph{Left panel:} The cumulative distribution of period ratios for adjacent planets.
The gray curve shows the observed distribution.
The green and red curves are for simulated systems at the disappearance of the disk of gas, before their dynamical instability.
The brown and blue curves show the distribution after the dynamical instability.
The green and brown distributions are from simulations which included the effects of turbulence in the disk of gas, and are essentially identical to those (red, blue) not accounting for turbulence.
\emph{Center panel:} The cumulative distribution of separations of adjacent planets in units of mutual Hill radii $R_H$.
The gray curve shows the observed distribution and the blue, green and red curves show the distributions obtained after the instability, broken-down by planet mass-range.
\emph{Right panel:} Normalized histogram of number of planets detected by transit in a given system.
The green histogram depicts the distribution observed by Kepler; the blue histogram shows the distribution expected for systems that underwent instability and the red histogram that of system that did not evolve in the aftermath of gas-removal.
The grey histogram fitting the observed distribution is obtained assuming that 90\% of observed system underwent instability.
Adapted from \citep{Izidoro2017}.\label{fig:1} }
\end{figure}

A second line of evidence for instabilities is that more than 50\% of the observed systems of transiting super-Earths are made of a single planet.
Either these planets are truly alone, which would be surprising, or there must be enough dispersion in the inclinations of the planets that only one of them is transiting \citep{Fang2012,Johansen2012,Zhu2020}. %
The inclination damping in the gas-disk should produce co-planar systems \citep{Cresswell2008}, so the inclination dispersion needs to have been acquired after gas-dispersal.

\citet{Izidoro2017,Izidoro2021} showed that the observed distribution of orbital \hbindex{period ratios} between adjacent planets (Fig.~\ref{fig:1}, left), as well and that of orbital separation in terms of mutual Hill radius (Fig.~\ref{fig:1}, center), are very well reproduced by simulated systems that are in resonant chains at the time the gas is removed from the system and that become unstable afterward.
The systems that experience this post-gas instability also reproduce the statistics of the number of planets detected by the Kepler telescope around individual stars (Fig.~\ref{fig:1}, right).
Finally, \cite{Millholland2021} found that the rarity of transit duration variations even in single-transiting systems is most consistent with a continuous distribution of mutual inclinations, rather than two distinct populations.
This unimodal but broad distribution is the expected outcome of systems that underwent dynamical instabilities of varying intensities \citep{He2020}.
Thus, post-gas dynamical instabilities should have been the norm in super-Earth systems.

A final indication for evolution of planetary systems after gas dissipation comes from the observations of debris disks with hot dust around main sequence stars, such as $\eta$~Corvi \citep{Lisse2012,Marino2017} or Vega \citep{Marboeuf2016}.
The dust produced in the vicinity of these stars cannot be sustained by the collisional grinding of local planetesimal populations because these populations could not remain massive for long enough.
Instead, the best explanation is that planetesimals from a distant disk are scattered inward as comets by planets \citep{Wyatt2007}.
The scattering of a large number of planetesimals for a long time requires that planets are currently migrating through the planetesimal disk due to the scattering process itself \citep{Bonsor2014}.
Beyond the evidence for violent dynamical reorganization, some observations point toward more quiet evolution paths occurring over the lifetime of the host star. In particular, close-in planets are heavily influenced by tidal interaction with the star, leading to gradual orbital circularization and decay.

Prompted by all these observational indications that evolution of planetary systems in the aftermath of gas removal is far from trivial, the next sections will examine the processes causing this evolution and their consequences.

\section{Conditions, timescales and outcome of planetary system instabilities}
\label{sec:selfinst}

Determining the conditions under which planetary systems remain stable is a centuries-old question in celestial mechanics (see \citealp{Laskar2013} for a historical review on the Solar System stability question).
Stable configurations are found if \hbindex{close encounters} between planets are not possible.
Indeed, strong planet interactions occur when their mutual interaction dominates the attraction of the star.
In this case, the planets exchange large amount of angular momentum and orbital energy, leading to collisions or planet ejections.
For non-crossing orbits, the planet interactions are not strong enough to significantly modify their orbits, and the planets evolve on Keplerian orbits that are slowly deformed over time.

An important feature of planet dynamics emerges if the motion of the planets can be averaged over their orbits.
In the averaged system (the dynamics are then called {secular}), the semi-major axes are constant and the dynamics only affect the planet mutual inclinations and eccentricities degree of freedom.
Combined with the conservation of the total angular momentum, the conservation of the semi-major axes in the secular problem leads to the conservation of a quantity called the total \hbindex{Angular Momentum Deficit} \citep[AMD,][]{Laskar1997}
\begin{equation}
    \mathrm{AMD} = \sum_k m_k\sqrt{\mathcal{G}M_\star a_k}\left(1-\sqrt{1-e_k^2}\cos(i_k)\right),
\end{equation}
where $m_k,a_k,e_k,i_k$ are respectively the mass, semi-major axis, eccentricity and inctlination of the $k$-th planet, $\mathcal{G}$ is the gravitational constant and $M_\star$ is the stellar mass.
The AMD bounds the eccentricities and, as a result, restricts the planets' excursions. A low enough AMD can forbid planet close encounters and ensure a system long term stability.
This form of stability, called AMD-stability \citep{Laskar2017}, gives a sufficient condition for the stability of a system, provided that the secular averaging is possible.
Most of the observed exoplanet systems with well characterized eccentricities are AMD-stable. 
Systems composed of largely unequal-sized planets such as the Solar System typically do not verify this condition. Indeed, the large AMD of the outer planets allows for large increase of the inner planet eccentricities, with collisions becoming possible if such configurations are reached.
Over the 5 Gyr of lifetime of the Sun, this scenario only occurs in roughly 1\% of the cases \citep{Laskar2009}.

Yet, it is quite common that the systems emerging from the protoplanetary disk cannot be considered secular since the averaging over the planet orbit is not a valid assumption in the presence of mean motion resonances.
As a result, these systems can still undergo a phase of dynamical rearrangement early in their lifetime.
The onset of stable configurations must therefore be amended to also identify the regions where the semi-major axes are not constant.

\subsection{Compact system instability}

A first analysis of stability of two adjacent planets on circular orbits was done by \cite{Gladman1993} in the framework of the Hill problem.
He found that the planets are Hill-stable (i.e. stable against mutual close approaches for all times) if their orbital separation $a_2-a_1$ exceeds $2\sqrt{3}$ mutual Hill radii ($R_H$), where $R_H =[(m_1+m_2)/(3M_{\star})]^{1/3} (a_1+a_2)/2$, $m_1$ and $m_2$ are the masses of the two planets with semi-major axes $a_1$ and $a_2$, and $M_{\star}$ is the mass of the star.
This criterion is the translation to circular orbits of a general topological invariant that exists in the three-body problem \cite{Marchal1982} that can be expressed explicitly in terms of a critical AMD value for general orbits \citep{Petit2018}. 
Unstable systems lead to collisions or ejections over very short timescales (a few thousands orbits).

The dynamical source of the instability arises from the action of the \hbindex{mean motion resonances} (MMR).
At small orbital separation, MMRs become wider than their mutual separation; thus they overlap, generating large-scale chaos \citep{Chirikov1979}. This leads to the diffusion of the semi-major axes and eventually to the crossing of the orbits.
The criterion for overlap for the largest MMRs (i.e. MMRs of first-order where the period ratio is close to $k/(k-1)$) was first derived in two-planet systems by \cite{Wisdom1980} for circular orbits and generalized to eccentic configurations by \cite{Deck2013} and \cite{Hadden2018}.
This instability threshold remains close to the Hill stability limit but shows a different scaling in the planet to star mass ratio $\varepsilon = (m_{\rm pl}/m_\star)$ as the minimal separation scales as $\varepsilon^{2/7}$ for circular orbits and the generic criterion scales with $\varepsilon^{1/4}$.

\begin{figure}[t]
    \centering
    \includegraphics[width=0.8\linewidth]{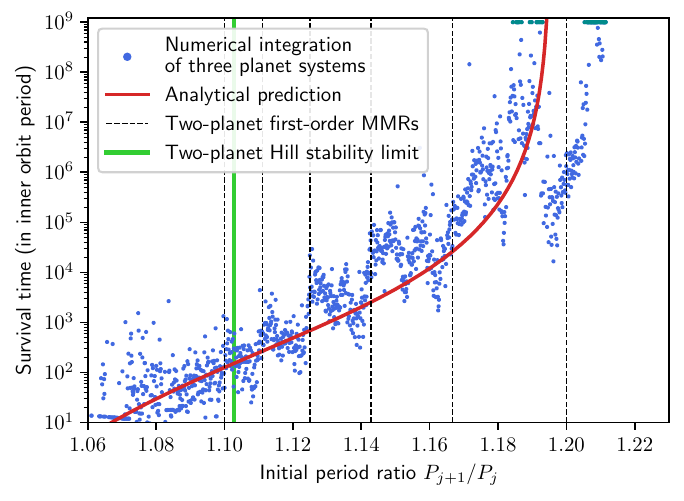} 
    \caption{The instability time as a function of mutual separation between adjacent planets in a three $3\, \mathrm{M_\oplus}$ mass planet systems (adapted from \citealp{Petit2020}). 
    The red line is the analytical instability time estimated from the diffusion of the semi-major axes due to three-planet zeroth-order MMRs \citep[][Eq. 81]{Petit2020}.
    Systems with a different taint of blue were stable for one billion orbits.
    The vertical dashed lines show the two-planet first-order MMR that modulate the trend in instability times and the green line indicates the Hill stability for a pair of planets \citep{Gladman1993}.}
    \label{fig:instabilitytimes}       %
    \end{figure}

The sharp stability boundary has no equivalent in systems with more than three planets.
\cite{Chambers1996} found numerically that multiplanetary \hbindex{compact systems} experience a long quiescent phase before a very rapid transition to collisional dynamics.
The instability timescale also scales exponentially with the spacing between the planets.
\cite{Chambers1996} proposed to scale the spacing with the mutual \hbindex{Hill radius} of the planets,
\begin{equation}\label{eq:Chambers1996_Inst}
    \log T_\mathrm{ins} \sim  b \Delta +c\ ,
\end{equation}
where $\Delta$ is the mutual separation in units of $R_H$ and $b$ and $c$ are constants.
They however showed that a scaling as $\varepsilon^{1/4}$ is more accurate over a large mass ratio regime.
Many works further confirmed that pioneering result \citep[\emph{e.g.}][]{Faber2007,Smith2009} but most authors continued using the Hill radius to rescale the problem.
For practical stability over stellar lifetime, separations over roughly 10 mutual Hill radii give a good stability boundary for super-Earth sized planets.
On top of the exponential trend, \cite{Obertas2017} revealed distinct modulations superimposed on this relationship in the vicinity of first and second-order mean motion resonances (corresponding to period ratios $k/(k-1)$ and $k/(k-2)$, for integer $k$, respectively) of adjacent planets (Fig.~\ref{fig:instabilitytimes}).
Chaotic diffusion induces a scatter of the instability timescales \citep{Hussain2020}.

The source of this wide variation in the \hbindex{instability timescale} was first proposed by \cite{Quillen2011} who suggested that the overlap of three-planet resonances (similar to the Laplace resonance between the Galilean moons) leads to a slow diffusion of the semi-major axes until a pair of planets encounter a strong two-body MMR which brings the orbits to crossing configurations.
While they correctly identified the source of instability, the model lacked predictive power.
\cite{Petit2020} refined the result by providing the first model determining the criterion for the three-body MMR overlap and a quantitative description of the diffusion process.
They inferred the timescale for instability for arbitrary planet masses and orbital spacing.
The analytical model does not predict an exponential growth with distance. Instead, it predicts that beyond a critical separation, the resonances no longer overlap, leading to much a greater survival time for the system (see Figure \ref{fig:instabilitytimes}).
This feature is actually visible on simulations run for more than 10 billion orbits \citep{Obertas2017}.
Furthermore, in this model, the critical stability separation scales as $\varepsilon^{1/4}$, which is consistent with numerical simulations.
The imprint of this stability limit is also visible in the observed exoplanet population:
giant planets are observed with smaller mutual separations if measured in Hill radii than super-Earths, whereas rescaling the separation by the $\varepsilon^{1/4}$ overlap criterion shows no mass dependence \citep{Petit2022}.

The case of eccentric orbits was studied by combining the approach of \cite{Petit2020} with the eccentric two-planet overlap criterion \citep{Hadden2018} by \cite{Tamayo2021}.
The three-body resonance network is enhanced at the vicinity of the main two-planet resonances as shown by \cite{Rath2022} and \cite{Lammers2024}.

\subsection{Resonant systems instability}

The picture of stability changes considerably in the case of systems with active \hbindex{mean motion resonances}.
The expectation for multi-resonant configurations comes from the realization that many super-Earths and sub-Neptunes could only have survived the gas disk phase if they were stopped from migrating all the way into the central star.
\hbindex{Type I migration} is suppressed at locations of the disk where density or temperature gradients are particularly steep \citep{Masset2006}, which could correspond for example to the outer radius of a magnetic cavity \citep{Ferreira1995,Armitage2020} or to the transition between an MRI active, low-density region of the disk and a low-viscosity, high-density region \citep{Gammie1996,Fromang2002,Hasegawa2010,Hasegawa2011,Bitsch2015}.
In these conditions a system of migrating planets necessarily experiences convergent migration from the moment the first of them reaches the planet trap, providing the possibility to capture in mutual mean motion resonances \citep{Batygin2015,Kajtazi2023}.
This is expected regardless of whether the super-Earths grow in the inner disk \citep{Ogihara2015} or migrate from beyond the snowline \citep{Cossou2014}.

As mentioned earlier, the dearth of resonant systems in the sample indicates that such systems are prone to instabilities after the disappearance of the disc. For the case of first-order mean motion resonances assembled during the disc phase, numerical results on stability thresholds after gas removal have been obtained by \cite{Matsumoto2012} and \cite{Matsumoto2020}. 
For a given resonant chain (i.e.,\ a given index $k$ of a $k$:$k-1$ resonance), \cite{Matsumoto2012} found a critical number of planets beyond which stability is not possible. This critical number $N_\mathrm{crit}$ depends on the resonant configuration and the planetary masses: at equal planetary masses, more compact resonant chains (higher $k$, e.g.\ each planet being in the 8:7 resonance with its neighbour) have smaller $N_\mathrm{crit}$ than more spaced-out resonant chains (lower $k$, e.g.\ in 6:5 resonances), and more massive planets have a smaller $N_\mathrm{crit}$ than less massive planets. When $N>N_\mathrm{crit}$, the instability time is similar to the non-resonant case (i.e.\ formula (\ref{eq:Chambers1996_Inst})).
\cite{Matsumoto2020} investigated the effect of the long-term mass evolution of planets and stars (driven by atmospheric escape and solar wind or coronal mass ejections respectively). Within plausible mass loss fractions (10\% for planetary atmospheric escape and 1\% for the host star), they found that systems with $N\lesssim N_\mathrm{crit}$ can also be destabilized. For example, for a 4:3 resonance chain with $m_\mathrm{pl}=10^{-4} M_*$, systems of up to $N=N_\mathrm{crit,4:3}=6$ planets always remain stable, but they become unstable after a planetary mass loss of about 10\% or a stellar mass loss of about 1\%. 

{
These numerical results can be understood on dynamical grounds \citep{Pichierri2018,Pichierri2020,Goldberg2022}. \cite{Pichierri2018} find that, in the case of two planets in a first-order resonance,{ the instability occurs when the masses are large enough that their mutual Hill radius becomes larger than $\sim 1.3$ times the minimal approach distance that the planets have from each other, if they are placed at the center of the resonance. This factor of $1.3$ is much larger than the one ($1/(2\sqrt{3})\sim 0.3$) for non-resonant orbits, but it is found to decrease with the amplitude of libration of the resonant mode. For the general case of multiple planets in a resonant chain, \cite{Pichierri2020} find that a richer dynamics is at play. In this case, the instability is triggered when a combination of libration frequencies of the mean motion resonant modes along the chain resonates with an appropriate synodic frequency. When a secondary resonance of this type is crossed, the system is excited away from the mean motion resonance libration centre and can thus be destabilised by close encounters. The resonant libration frequencies scale with a power of the planetary mass, while the slowest synodic frequency only depends on the number $N$ of planets and the resonance configuration and in particular decreases with $N$ and how compact the chain is. Combining these two facts, the stability criterion based on the crossing of these double resonances yields a qualitative explanation to the numerical findings of \cite{Matsumoto2012}, \cite{Matsumoto2020}. These ideas have been further carried out in \cite{Goldberg2022}, using analytical estimates of the resonant and synodic frequencies to obtain a quantitative criterion for the critical mass above which a resonant system goes unstable. This simple criterion accurately predicts the stability limits of resonant chains in systems composed of up to six planets without external perturbations, after the disappearance of the gaseous disk.} 
  
\cite{Izidoro2017} find that only $<10\%$ of the \hbindex{resonant chains} assembled via disc-driven migration can remain stable in order to match the period distribution of observed exoplanets. It is not clear whether such high ($>$90\%) instability fractions can be reached by post-gas instabilities unaided by external perturbations or other physical processes. \cite{Izidoro2017} found that only 50\% of the systems assembled via disk-driven migration in their simulations become unstable by themselves after gas removal, while \cite{Izidoro2021} found that high instability fractions can be obtained, but only in models that produce more massive planets. 
This is consistent with the theoretical results mentioned above, but the needed planetary masses are higher than the ones observed in the Kepler sample.
Thus, the effect of external perturbations might play a significant dynamical role in determining the fate of these systems, as will be discussed in the next section.

\subsection{Dynamical reorganization of unstable systems}

Once a planetary system becomes unstable, in absence of damping exerted by the gas or planetesimal dynamical friction, the fate of the system depends on the so-called \hbindex{Safronov number}.
This is 
\begin{equation}
\Theta=\frac{v_\mathrm{esc}^2}{2v_\mathrm{orb}^2} \ ,
\end{equation}
 where $v_\mathrm{esc}$ is the escape velocity from the planets and $v_\mathrm{orb}$ is their orbital velocity on a circular orbit \citep{Safronov1972}.
If this number is larger than one, the close encounters among the planets are likely to lead to the \hbindex{ejection} of some of the bodies until a new stable configuration is achieved.
In fact, mutual scattering tends to give the planets a velocity dispersion of order $v_\mathrm{esc}$.
If the latter exceeds the escape velocity from the potential well of the star, which is equal to $\sqrt{2} v_\mathrm{orb}$, an ejection from the system is inevitable.
If instead the Safronov number is smaller than unity, \hbindex{collisions} among the planets are more likely.
In this case, the reduced number of planets (albeit with larger masses due to the merging collisions) also eventually leads to a new stable configuration.

In the case of giant planets in the outer part of the disk, $\Theta$ is typically larger than~1 and therefore planets are ejected when instability occurs.
This was probably the fate of extrasolar giant planet systems, leaving behind a single detectable planet on eccentric orbit \citep{Ford2008,Juric2008,Chatterjee2008}.
For close-in super-Earth systems, instead, $\Theta<1$ and, therefore, planets merge in collisions when they become unstable \citep{Izidoro2017}.
An important point is that collisions may happen before the velocity dispersion has the time to reach the asymptotic value $v_{\rm esc}$, so that, when the number of planets decreases enough to achieve a stable configuration, the system can remain frozen with an eccentricity and inclination excitation smaller than expected \citep{Esteves2020}.
Kepler-11 \citep{Lissauer2011} is such an example of these nearly coplanar and circular non-resonant systems.

The classic model of formation of terrestrial planets \citep{Chambers1998,Agnor1999,Chambers2001} in the Solar System is based on the self-instability of the inner solar system planetary embryos after gas-removal.
In fact, the planetary embryos formed in the inner part of the protoplanetary disk by oligarchic growth are expected to have been approximately Mars-mass and separated by 5-10 mutual Hill radii \citep{Kokubo2000}.
Due to their large number and short spacing, the system of planetary embryos becomes naturally unstable on a short timescale after gas removal.
Exoplanets may follow a similar formation scenario by the accretion of planetesimals from a narrow ring around one AU, where close encounters efficiently lead to collisions and fast growth if the planetesimal reservoir is large enough \citep{Batygin2023}.

\section{Evolution of planetary systems under external perturbations} 
\label{sec:external}

Beyond the evolution purely driven by the planetary dynamics, many external processes reshape planetary systems and can eventually lead to system reorganization. These events can be violent, with phases of dynamical instabilities, or completely quiescent. This section describes various mechanisms taking place during the main lifetime of exoplanetary systems, after the \hbindex{protoplanetary disk} has been dissipated.

\subsection{Interactions with the disappearing gas-disk}
\label{sec:gas}

Turbulence in the disk may prevent capture of the planets deeply into resonances \citep{Adams2008,Batygin2017,Hands2018}, favoring their eventual orbital instability.
As it was said before, instabilities within the disk of gas are likely to be erased because the damping action of the disk will quickly recapture planets into resonance \citep{Lega2013}.
Thus, for turbulence to be a trigger of an instability that can leave a permanent imprint on the final orbital structure of the system, the disk has to remain strongly turbulent until it disappears.
It is even possible that disks become more turbulent near the end of their lifetime.
As the disk gas gets depleted, the ionization of gas in the midplane becomes possible, potentially activating the magneto-rotational instability, enhancing the turbulence.
However, if the density of gas is weak, the effects of turbulence may be insufficient to extract the planets from the core of mean motion resonances where they are likely to have been captured when the disk was laminar in the midplane \citep{Deck2015}. \cite{Izidoro2017} modeled disk turbulence 
and concluded that turbulence does not enhance the probability that a system of super-Earths becomes unstable.
This result, however, may depend on the assumed scaling of turbulence strength with orbital radius.

As the disk disappears, the \hbindex{magnetic cavity} is likely to expand because the balance between the magnetic torque and the viscous torque (that sets the disk's truncation) occurs farther and farther from the star as the disk's density is reduced \citep{Ferreira1995,Armitage2020}.
\citet{Liu2017} considered the effects that the expansion of the magnetic cavity has on a system of resonant super-Earths near the disk's inner edge.
They find that if the expansion of the cavity is slow enough and the disk beyond the cavity is still massive enough compared to the planets, the planet at the disk's edge remains locked with the edge and migrates outwards with it as the cavity expands.
 Eventually, the inner planet can no longer follow the radial motion of the disk's edge and ends up in the cavity, where its migration stops.
The receding edge can then entrain the second planet outwards for a while, and so forth.
The result is that the planets are extracted from their original resonance and deposited on quasi-circular and co-planar orbits with wider and non-resonant orbital periods.
Even if this can explain the lack of preference for resonant ratios seen in the Kepler data, this process alone cannot fully explain the data: for instance, the fraction of stars with multiple transiting planets would be too large because of the coplanar geometry of the final planetary systems \citep{Izidoro2017}.
However, as planets depart from their original resonance, they can cross new resonances which can excite their orbital eccentricities enough to trigger orbital instabilities \citep{Liu2022}.
How frequently an instability can occur in this process has not been quantified. In the case of the TRAPPIST-1 system, this process can explain through a quiescent migration the high-order 8:5 -- 5:3 resonances observed in the inner system, starting from an originally more compact 3:2 chain \citep{Pichierri2024}.

Another changing property of a disappearing disk is its vertical aspect ratio.
In the inner part of the disk the dominant heating mechanism which sets the aspect ratio is viscous heating, which decreases with decreasing accretion rate onto the central star \citep{Bitsch2015}.
As the aspect ratio of the disk decreases, the eccentricity of resonant planets undergoing damping from the disk also decreases (see e.g.\ \citealt{Xu2017}).
With decreasing eccentricity the frequency of libration of resonant planets increases \citep{Batygin2013}.
This may cause instabilities due to the passage through commensurabilities among the libration periods of multiple resonant planets or between libration and synodic periods.

\subsection{Interaction with remnant planetesimals}
\label{sec:planetesimals}

The removal of the gas leaves behind unaccreted \hbindex{planetesimals}.
The interaction between planets and planetesimals can be neglected as long as there is sufficient gas in the system, but it becomes important once the gas disk is substantially depleted \citep{Capobianco2011}.
The scattering of planetesimals by a planet changes the orbit of the planet by the action-reaction principle.
That is, a single planet embedded in a planetesimal disk typically migrates inwards, because of a scattering bias \citep{Kirsh2009} that favors the scattering of planetesimals in the outwards direction.
However, two (or more) planets on nearby orbits typically migrate in divergent directions, the outer planet(s) moving outwards and the inner one inwards, because the outer planet acts as a conveyor belt, transferring planetesimals from the outer disk to the inner planet (\citealp{Fernandez1984}; see \citealp{Levison2007} for a review of planetesimal-driven migration).

In the most recent formulations of the \hbindex{Nice model}, the giant planets are initially in a mean-motion resonant chain.
Scattering of planetesimals by the planets extracts the planets from their original resonances through divergent migration.
The crossing of other higher-order resonances during this evolution gives the planets enough eccentricity excitation to trigger a global instability \citep{Morbidelli2007}.
If the planetesimal disk is far enough so that the planets cannot scatter them, the secular planet-planetesimal interactions can still modify the resonant orbits of the planets and enhance their eccentricities until a global instability follows \citep{Levison2011}.
However, the flux of dust generated by the distant planetesimal disk due to its slow collisional grinding can also drive divergent migration of the planets \citep{Deienno2017}.
During the instability phase, most of the eccentricity and inclination excitation and eventual extraction from resonance is caused by mutual close encounters between the planets themselves, which are more violent than scattering of planetesimals.
The planetesimals nevertheless still play a fundamental role: by exerting dynamical friction on the planets, they eventually damp the planetary eccentricities and inclinations, allowing the planets to recover a stable configuration with moderately excited orbits \citep{Tsiganis2005,Morbidelli2007,Nesvorny2012}.
In this process, the planetesimals are violently dispersed and only those landing in stable niches of the orbital space survive.
Thus, reproducing the current orbital structure of the surviving planetesimal populations is a crucial diagnostic of an instability model, possibly more than the final orbits of the planets themselves.
Thanks to the analysis of the different minor bodies reservoirs (main asteroid belt, Jupiter Trojans, irregular satellites, Kuiper Belt) in the Solar System, works based on the Nice model have enabled the development of a comprehensive scenario for its dynamical history \citep{Nesvorny2013,Morbidelli2010,Roig2015,Deienno2016,Nesvorny2015,Nesvorny2015a}.
This approach has also started to be used to interpret exoplanet systems where debris disks are observed.
For instance, \cite{Beust2024} concluded that the two planets in the $\beta$~Pictoris system are compatible with the presence of a reservoir around 1.5 AU from which originate the observed exocomets transiting the star.

\citet{Chatterjee2015} proposed that divergent migration due to planetesimal scattering is also the dominant process that extracts super-Earths from their original resonant chain.
As noted above (concerning the cavity expansion mechanism of \citet{Liu2017}), simply extracting the planets from the resonances on circular and co-planar orbit would not be sufficient to explain the observations.
\citet{Raymond2022} found that a few remnant planetesimals can also cause the disruption of a long resonant chain such as Trappist-1, which emphasizes the fragility of such configurations.
A dynamical instability is needed to produce the inclination excitation deduced from the frequency of multiple transiting planets \citep{Izidoro2017}.
However, it is unclear what the distribution and the total mass of planetesimals in the vicinity of super-Earths are at the disappearance of the gas disk.
If the super-Earths migrated from larger distances, it is likely that their broad neighborhood was substantially depleted of planetesimals during the migration phase.
If this is true, the remnant planetesimals would not carry enough mass to be able to substantially change the planets' orbits in the aftermath of gas removal.

\subsection{Tidal interactions}
\label{sec:tides}
Planets on orbits close to the central star may undergo substantial orbital evolution due to \hbindex{tidal dissipation}.
Dissipation can occur both in the star, which primarily drives the orbital migration of the planet, and in the planet, which primarily damps its orbital eccentricity \citep[][see also the chapter by Mathis]{Goldreich1966}. 

Orbital migration due to dissipation within the star is probably only relevant for the shortest period hot Jupiters, which are capable of raising a significant tidal bulge on the star. 
The mechanism is the same as what drives the expansion of the Moon's orbit. The planet raises a tidal bulge on the star, but due to frictional dissipation the bulge points slightly ahead or behind of the planet rather than directly at it. 
The bulge thus imparts a torque on the planet and angular momentum is exchanged between the stellar rotation and planetary orbit.
The direction of migration depends on the relative frequencies of the stellar rotation and planetary orbit. Migration is inward for planets whose orbital periods are less than the stellar rotation period, the situation expected for most hot Jupiters around main sequence or evolved stars.

Inward orbital decay has been conclusively detected for WASP-12 b, a hot Jupiter with a period of 1.09d 
 \citep{Patra2017,Yee2020}. 
In fact, WASP-12 b is expected to cross the Roche limit within several Myr, leading to bulk mass loss from the planet or its complete disintegration. Tidal dissipation can therefore be a mechanism to remove planets entirely. Indeed, the population of close-in giant planets appears to be shaped by this effect \citep{Matsakos2016,Hamer2019}.

Dissipation within the planet is a closely related physical process but tends to have different implications for the evolution of the planetary orbit. 
Because close-in planets are likely predominantly tidally locked, angular momentum transfer is negligible and the dominant effect is a decrease in the orbital eccentricity. 
The shortest period planets are therefore expected to have mostly circular orbits, even if they were delivered to their current orbit via a high-eccentricity mechanism. 
This plays an important role in a number of proposed formation mechanisms of hot Jupiters and small \hbindex{ultra-short-period planets} \citep{Fabrycky2007,Beauge2012,Pu2019}.

Damping in eccentricity becomes especially relevant and complex for pairs of planets near first-order mean-motion resonances.
In that case, the eccentricity damping forces the planets to migrate away from each other, due to the shape of the resonant locus in the eccentricity vs. semi-major axis plane \citep{Papaloizou2010,Lithwick2012,Batygin2012}. 
Thus, the period ratios of pairs of near-resonant planets can gradually increase 
over the lifetime of the system, an effect termed `resonant repulsion.'
Although the magnitude of the increase is small, transiting planets have precisely measured orbital periods, and in fact the distribution of period ratios of adjacent planets shows a clear excess of planet pairs $\sim 1\%$ wide of, and a paucity narrow of, some first-order mean-motion resonances \citep{Fabrycky2014}. 
\hbindex{Tides} have been invoked as an explanation of the shift in the period ratio distribution \citep{Lithwick2012,Lee2013}, and can reproduce this pattern if typical values of the tidal quality factor $Q$ are comparable to that of the Earth, somewhat stronger than what is expected for dry rocky planets (however, tides may not fully account for the eccentricity distribution, see below). 

In systems of three or more planets, resonant repulsion continues to operate, with the additional consequence that the three-body commensurability generated by the difference of two-body resonances persists during the repulsion \citep{Papaloizou2015}. 
For example, a trio of planets in pairwise 3:2 resonances will initially satisfy $2n_1-3n_2\approx 0$ and $2n_2-3n_3\approx 0$, where $n_i$ is the mean motion of the $i$-th planet. 
As resonant repulsion proceeds, each of these quantities will increase away from zero, but at precisely the same rate, so that $(2n_1-3n_2)-(2n_2-3n_3)\approx 0$ still holds. Consequently, the three-body Laplace angle, $\phi_{3b}=2\lambda_1-5\lambda_2+3\lambda_3$, where $\lambda_i$ is the mean longitude of the $i$-th planet, will remain in libration.
This angle, being zeroth-order in eccentricity, can be readily estimated from transit data \citep{Siegel2021}, and is frequently found to be librating in near-resonant systems. 

Nevertheless, the precise nature and role of tidal interactions, particularly in small planet systems, are highly uncertain on both a theoretical and observational level. Theoretical predictions have been hampered by our near-ignorance of the \hbindex{tidal quality factor}, $Q$, which characterizes the strength of dissipation, for different bodies. Within the Solar System, measurements of the migration of natural satellites point to significant diversity in $Q$ attributable to differences in interior composition and structure \citep{Goldreich1966,Lainey2016}. Even when the interior structure is known, $Q$ can depend sensitively on the orbital period and predictive models are challenging \citep{Efroimsky2007,Ogilvie2004,Ma2021}. Adding further complication, tidal dissipation can dramatically strengthen if, during its tidal spindown, the planet is captured into a secular \hbindex{spin-orbit resonance} with nonzero spin obliquity \citep{Millholland2019,Su2022}.

Observationally, it is not clear whether trends in systems of multiple small planets are actually consistent with the impacts of tidal evolution. For one, tidal evolution should be much more pronounced in the closest-in planets, due to the steep dependence of the damping rate on orbital radius. While \cite{Delisle2014} claimed that this effect is apparent, \cite{Choksi2020} argue that the tides cannot be the sole driver of the period ratio distribution near resonance. Furthermore, planet pairs that gained their offsets from exact resonance due to tides should have extremely small eccentricities today, but measurements from transit timing variations indicate that these eccentricities are moderate \citep{Choksi2023,Goldberg2023}. Finally, working backward, it is possible to determine the initial eccentricity required to generate the current spacing. \cite{Silburt2015} demonstrated that in many cases, these eccentricities are so high as to make the system unstable. 

It is important to note that tidal dissipation is only one example of a dissipative process that removes orbital energy from the system. 
Other sources of damping, such as interactions with the gaseous disk or a dynamical friction with a belt of planetesimals described above, can have a similar impact on the planetary orbits. 
For example, \cite{Papaloizou2018} proposed that the inner 8:5--5:3 triple in TRAPPIST-1, not an expected outcome of Type I migration, could be produced by tidal damping away from an initial 3:2--3:2 chain with preservation of their Laplace angle.
However, in this model not all 7-planets are linked by \hbindex{Laplace resonances}, which is the currently observed state \citep{Agol2021}. With the updated orbital characterization, \cite{Brasser2022} demonstrate that such tidal evolution applied to the full chain would result in an incorrect configuration of the outer planets.
As alternative explanations, mechanisms such as one-sided Lindblad torques \citep{Huang2022} and the recession of the inner edge of the protoplanetary disk \citep{Pichierri2024} have been proposed to spread the inner trio from an initial 3:2--3:2 chain into the 8:5--5:3 configuration before the outer planets arrive. 
More generally, the failure in some cases of basic tidal models to satisfy stricter observational constraints seems to point to a broader range of dissipative processes which operate over a range of timescales and with slightly different consequences. 

\subsection{Interactions with a stellar companion}
\label{sec:stellarcompanions}

The most well-known consequence of the interaction with a stellar companion is the so-called \hbindex{von Zeipel-Lidov-Kozai effect} (ZLK, \citealp{vonZeipel1910,Lidov1962,Kozai1962,Ito2019}, see \citealp{Naoz2016} for a thorough review).
A planet initially on a circular orbit, perturbed by an inclined distant star will increase its orbital eccentricity while its orbital plane approaches that of the stellar companion.
If the orbit of the stellar companion is circular, this process is reversible, leading to coupled oscillations of the eccentricity and inclination of the planet's orbit, driven by the precession of the orbit's argument of perihelion. 
The reversibility, however, can be broken by the tidal interaction with the central star.
As the ZLK oscillation brings the planet onto a high-eccentricity orbit, tidal circularization leads to a so-called high-eccentricity migration which has been suggested as responsible for the origin of the hot-Jupiters \citep{Fabrycky2007}.
As tides tend to damp efficiently the eccentricity but not the inclination, the planets following this route would tend to exhibit large obliquities \citep{Attia2021}.
Indeed, obliquity measurements of \hbindex{hot-Jupiters} suggests that a significant fraction of the population shows evidence of such past evolution \citep{Albrecht2021,Bourrier2023}.
Moreover, if the stellar companion has an eccentric orbit, the evolution of the planet is chaotic and the orbit can flip and become retrograde relative to the central star \citep[e.g.][]{Li2014}.

It should be stressed that the ZLK effect can be quenched if the  precession rate of the argument of perihelion induced by a disk, a planet companion, or general relativity is faster than that induced by the stellar companion \citep{Boue2014}.
Because the stellar companion, although more massive, is typically much farther away, mutual perturbations among planets in a system can easily dominate over the stellar perturbation.
A clear example of this is provided by the natural satellites of Uranus.
These satellites orbit on the planet's equatorial plane, and therefore their orbital inclination relative to the orbit of the Sun is 98 degrees.
The orbit of a single satellite would therefore be unstable to solar perturbations through the ZLK effect, but the satellite system as a whole is stable, thanks to the fast precession of the satellites' perihelia induced by mutual perturbations.
Similarly, \citet{Batygin2011} showed that a planet embedded in a disk or a system of two giant planets orbiting the central star on coplanar orbits, perturbed by a distant and inclined stellar companion, is likely to be stable against this external perturbation.
However, after the disk is removed, if the planets become unstable for some other reason and one of the two planets is removed, the remaining planet suddenly starts large eccentricity and inclination oscillations due to the ZLK effect.
In this case, the ZLK effect is the consequence of the instability in a planetary system, rather than the cause.

\section{Conclusions}
\label{sec:conclusions}

Observations suggest that the evolution of a planetary system is not finished when the gas of the protoplanetary disk is removed.
The \hbindex{architecture of planetary systems} can change profoundly; collisions between planets are possible if the planetary system becomes unstable and the Safronov number is smaller than unity.
Both the Solar System and most extrasolar planetary systems have been sculpted by post-gas evolution and planet instabilities.

The cause of these instabilities is still unclear as many mechanisms are viable.
Planetary systems which are too tightly packed become unstable all by themselves \citep{Chambers1996,Petit2020}, without the need of external perturbations.
However, recent numerical simulations show that only $\sim$50\% of the systems of super-Earths generated by gas-driven migration would become unstable this way; instead, fitting the observed distribution requires that the number of systems undergoing post-gas instability exceeds 90\% \citep{Izidoro2017}.
Thus, external perturbations should play a role in triggering \hbindex{instabilities} at sufficiently high rates.
There is probably no universal cause for instability. Several mechanisms can be at play and their relative importance likely differs on a case-to-case basis.
The interaction with a remnant planetesimal disk is perhaps the most generic of the processes, and it is the one responsible for the past instability of the giant planets of the Solar System.
However, it is not obvious that massive planetesimal populations could have survived at the end of the gas-disk phase if the planets had undergone large-scale migration within the disk.

Some progress might be possible with statistics of exoplanetary system architectures for stars in different environments (e.g.
binaries, in clusters or isolated) to highlight the importance of external perturbations.
But, most likely, significant progress will occur only after the development a robust understanding of the population of planets around young stars ($\lesssim 100$ Myr), or even better, those still embedded in their birth disks.
The results presented in this chapter are based on reconstructing the past evolution of mature planetary systems with necessarily uncertain modeling efforts.
The observation of planets around young stars will instead provide direct information on how and on which timescale planetary systems evolve.

\bibliographystyle{spbasicHBexo}
\bibliography{exohand.bib}

\section{Cross-references}
\begin{itemize}
    \item A Brief Overview of Planet Formation
    \item Planetary Migration in Protoplanetary Disks
    \item Formation of Giant Planets
    \item Formation of Super-Earths
    \item Tightly Packed Planetary Systems
    \item Tidal Star-Planet Interactions: A Stellar and Planetary Perspective
    \item Debris Disks: Probing Planet Formation
    \item TRAPPIST-1 and its compact system of temperate rocky planets
    \item WASP-12b: A Mass-Losing Extremely Hot Jupiter
\end{itemize}

\end{document}